\documentclass[useAMS,usenatbib]{mn2e}

\usepackage{graphicx}
\usepackage{epsfig,floatflt}
\usepackage{times}

\title[Spectroscopic binarity of FN and V1344~Aql]
{Discovery of the spectroscopic binary nature of the
classical Cepheids FN~Aql and V1344~Aql}

\author[Szabados et al.]{L. Szabados$^{1}$,  
B. Cseh$^{2,3}$, 
J. Kov\'acs$^2$, 
B. Cs\'ak$^2$,
\'A. D\'ozsa$^2$,
Gy. M. Szab\'o$^{1,2}$,
\newauthor
A.~E. Simon$^{1,2}$,
T. Borkovits$^{1,2,4}$,
L. L. Kiss$^{1,2,5}$,
I. Jankovics$^2$,
Gy. Mez\H{o}$^{1}$
\\
$^1$Konkoly Observatory, Research Centre for Astronomy and Earth 
Sciences, Hungarian Academy of Sciences, H-1121 Budapest, \\ 
Konkoly Thege Mikl\'os \'ut 15-17, Hungary\\
$^2$ELTE Gothard-Lend\"ulet Research Group, H-9704 Szombathely, 
Szent Imre herceg u. 112., Hungary\\
$^3$Department of Astronomy, Lor\'and E\"otv\"os University,
H-1117 Budapest, P\'azm\'any P. s\'et\'any, Hungary\\
$^4$Baja Astronomical Observatory, H-6500 Baja, Szegedi \'ut, Kt. 
766, Hungary\\
$^5$Sydney Institute for Astronomy, School of Physics, 
University of Sydney, NSW 2006, Australia\\
}

\begin{document}

\date{Accepted Received ; in original form }

\pagerange{\pageref{firstpage}--\pageref{lastpage}} \pubyear{2014}

\maketitle

\label{firstpage}

\begin{abstract}
We present the analysis of photometric and spectroscopic data 
of two classical Cepheids, FN~Aquilae and V1344~Aquilae. 
Based on the joint treatment of the new and earlier radial 
velocity data, both Galactic Cepheids have been found to be a 
member in a spectroscopic binary system. 

To match the phases of the earlier radial velocity data correctly
with the new ones, we also determined the temporal behaviour of the 
pulsation period of these Cepheids based on all available photometric 
data. The $O-C$ graph covering about half century shows slight changes 
in the pulsation period due to stellar evolution for both Cepheids.

\end{abstract}

\begin{keywords}
binaries: spectroscopic -- stars: variables: Cepheids -- stars: individual:
FN Aquilae -- stars: individual: V1344 Aquilae
\end{keywords}

\section{Introduction}
\label{intro}

Classical Cepheid variable stars are well known primary distance
indicators owing to the famous period--luminosity ($P$-$L$) 
relationship. Companions to Cepheids may complicate and at the
same time facilitate the applicability of using the $P$-$L$ 
relationship for distance determination. On the one hand, the 
photometric contribution of the secondary star has to be taken 
into account when determining the brightness and colours of the 
Cepheid component in optical photometric bands, 
otherwise the companion may falsify the
luminosity value determined for the Cepheid \citep{SzK12}. 
On the other hand, Cepheids in binary systems serve as reliable 
calibrators of the $P$-$L$ relationship \citep{E92}. The frequency 
of binaries among bright Cepheids exceeds 50\%, while among the 
fainter ones an observational selection effect encumbers revealing 
binarity \citep{Sz03b}.

In the case of pulsating variables, like Cepheids, spectroscopic
binarity (SB) manifests itself in a periodic variation of the
$\gamma$-velocity (i.e. the radial velocity, RV, of the mass 
centre of the Cepheid). In practice, the orbital RV variation of 
the Cepheid component is superimposed on the pulsational RV 
variations. An unrevealed orbital motion increases the scatter 
of the pulsational RV curve, and has an adverse effect on 
estimating the distance (therefore on the calibration of the 
$P$-$L$ relationship) via the use of the Baade--Wesselink method.

The orbital period of binaries involving a supergiant Cepheid 
component cannot be shorter than about a year. SBs involving 
a Cepheid component with orbital periods longer than a decade 
are also known (see the on-line data base on binaries among 
Galactic Cepheids:
http://www.konkoly.hu/CEP/orbit.html).
Therefore, a first epoch RV curve, especially based on data 
obtained in a single observational season, is usually
insufficient for pointing out an orbital effect superimposed
on the RV changes due to pulsation.

In this paper we point out SB of two bright Galactic Cepheids, 
FN~Aquilae (FN~Aql) and V1344~Aquilae (V1344~Aql) by analysing RV data. 
Basic information on these Cepheids is found in Table~\ref{obsprop}. 
The new RV data and the observational circumstances are described 
in Section~\ref{data}, then Sections~\ref{fnaql} and \ref{v1344aql} 
are devoted to the results on each new SB Cepheid, while 
Section~\ref{concl} contains our conclusions.

\section{New radial velocities}
\label{data}

Both Cepheids were observed among the targets of a RV survey of 
Galactic classical Cepheids initiated in 2012, using the 
0.5\,m RC telescope of ELTE Gothard Astrophysical Observatory, 
Szombathely and the 1\,m RCC telescope of Piszk\'estet\H{o} Mountain 
Station of the Konkoly Observatory of the Research Centre for Astronomy 
and Earth Sciences of the Hungarian Academy of Sciences. The spectrograph 
was the same fibre-fed instrument at both locations, the eShel system 
of the French Shelyak Instruments \citep{Thizy2011}. The detailed 
description of the instrument, methods and the observing programme 
is given in \citet{Csetal14}.

The exposure times were typically 900~sec, the observing sequence was 
ThAr-object-object-ThAr. In most cases two consecutive spectra were 
recorded, which were averaged at the end of reduction process.

All spectra were reduced with standard tasks in 
\textsc{iraf}\footnote{\textsc{iraf} is distributed by the National 
Optical Astronomy Observatories, which are operated by the Association 
of Universities for Research in Astronomy, Inc., under cooperative 
agreement with the National Science Foundation.}, including
bias, dark and flat-field corrections, aperture extraction, 
wavelength calibration, and continuum normalization. We checked 
the consistency of wavelength calibrations via RV standard star 
observations, which proved the stability of the system. 

RV values were determined by cross-correlating object
spectra and $\mathrm{R}=11500$ synthetic spectra chosen from the 
\citet{Munari2005} library using the FXCOR task of \textsc{iraf}. 
Correlations were calculated between 4870 and 6550~\AA, 
excluding Balmer lines, NaD and telluric regions.

Barycentric Julian dates and velocity corrections for mid-exposures 
were calculated using the \textsc{barcor} code of \citet{Hrudkova2006}.
This method resulted in a 100-200\,m\,s$^{-1}$ uncertainty 
in the individual RV values, while further tests have shown that 
our absolute velocity frame was stable to within 
$\pm$200-300\,m\,s$^{-1}$. 

The individual RV values are listed in Tables~\ref{tab-fnaql-data}
and~\ref{tab-v1344aql-data}, respectively.

\begin{table}  
\begin{center}  
\caption{Basic data of the programme stars 
and the number of spectra.} 
\label{obsprop}  
\begin{tabular}{|lccccc|} 
\hline  
Cepheid & $\langle V \rangle$ & P & Mode & Number & Number \\
& mag & (d)& of pulsation & of nights & of spectra \\
\hline  
FN~Aql    &  8.40 &  9.482 & Fundamental    & 17 & 22\\ 
V1344~Aql &  7.77 &  7.477 & First overtone & 38 & 73\\
\hline   
\end{tabular} 
\end{center}  
\end{table}

\begin{table}
\caption{New RV values of FN~Aql}
\begin{tabular}{lrc}
\hline
\noalign{\vskip 0.2mm}
JD$_{\odot}$ & $v_{\rm rad}$ & $\sigma$\  \\
2\,400\,000 + &(km\,s$^{-1}$) & (km\,s$^{-1}$)\\
\noalign{\vskip 0.2mm}
\hline
\noalign{\vskip 0.2mm}
56463.4765 & 18.78 & 0.12 \\
56464.4777 & 26.97 & 0.83 \\
56465.5075 & 31.13 & 0.11 \\
56467.4097 & 10.18 & 0.12 \\
56490.4477 &  5.26 & 0.28 \\
56491.4815 & 14.09 & 0.48 \\
56505.4206 &  9.57 & 0.23 \\
56506.4062 &  7.71 & 0.35 \\
56516.4299 &  5.85 & 0.16 \\
56519.5152 & 10.85 & 0.61 \\
56520.3778 & 18.96 & 0.15 \\
56521.4576 & 27.81 & 0.10 \\
56522.3047 & 31.66 & 0.12 \\
56523.3108 & 21.95 & 0.15 \\
56524.3069 &  9.95 & 0.11 \\
56570.3523 & 27.77 & 0.21 \\
56582.3024 &  7.42 & 0.26 \\
\noalign{\vskip 0.2mm}
\hline
\end{tabular}
\label{tab-fnaql-data}
\end{table}

\begin{table}
\caption{New RV values of V1344~Aql}
\begin{tabular}{lrc}
\hline
\noalign{\vskip 0.2mm}
JD$_{\odot}$ & $v_{\rm rad}$ & $\sigma$\  \\
2\,400\,000 + &(km\,s$^{-1}$) & (km\,s$^{-1}$)\\
\noalign{\vskip 0.2mm}
\hline
\noalign{\vskip 0.2mm}
56053.5336 &  2.89  & 0.10\\
56058.5486 & $-$7.18& 0.09\\
56059.5612 & $-$3.58& 0.09\\
56084.4737 & $-$0.96& 0.62\\
56091.4160 &  2.19  & 0.40\\
56104.4768 & $-$3.27& 0.48\\
56106.5345 &  1.84  & 0.20\\
56119.4578 & $-$3.05& 0.17\\
56122.3520 & $-$5.49& 0.10\\
56149.3418 & $-$3.08& 0.17\\
56151.4488 &  1.70  & 0.16\\
56152.3518 & $-$6.46& 0.19\\
56153.3440 &$-$11.97& 0.16\\
56154.3298 &$-$10.63& 0.18\\
56155.3708 & $-$8.05& 0.14\\
56157.3852 &  0.18  & 0.17\\
56159.4025 & $-$2.07& 0.25\\
56161.4092 &$-$11.42& 0.19\\
56163.3492 & $-$6.85& 0.11\\
56164.3200 & $-$3.41& 0.16\\
56165.3170 &  1.69  & 0.12\\
56172.4338 &  0.13  & 0.31\\
56463.4200 & $-$1.79& 0.52 \\
56464.4263 &  2.89  & 0.38 \\
56465.4884 &  2.59  & 0.12 \\
56467.3852 & $-$10.32& 0.30 \\
56490.3670 & $-$9.86 & 0.30 \\
56491.3978 & $-$7.47 & 0.29 \\
56504.3974 & $-$9.31 & 0.15 \\
56505.3915 & $-$9.88 & 0.18 \\
56506.3783 & $-$7.38 & 0.17 \\
56516.4160 &  2.17   & 0.11 \\
56520.3647 & $-$9.69 & 0.10 \\
56521.4696 & $-$7.19 & 0.09 \\
56522.3177 & $-$5.05 & 0.09 \\
56523.3230 & $-$0.99 & 0.11 \\
56570.3760 &  1.56   & 0.20 \\
56582.3282 & $-$4.14 & 0.15 \\
\noalign{\vskip 0.2mm}
\hline
\end{tabular}
\label{tab-v1344aql-data}
\end{table}

\section{Results for FN~Aquilae}
\label{fnaql}

\subsection{Accurate value of the pulsation period}
\label{fn-period}

Variability of FN~Aql was discovered by Cerasskaya in 1929
\citep{B29}. The Cepheid nature of the brightness variations
and the pulsation period of about 9.5~d was established by
Lause \citep{P31}. Cepheids pulsating with such a period 
have moderate amplitudes and a bump near the phase of maximum
brightness, and their oscillation corresponds to the fundamental
mode. 

To separate the orbital and pulsational effects in the RV
variations, knowledge of the accurate value of the pulsation 
period is essential, especially when comparing RV data obtained 
at widely differing epochs. Use of the accurate pulsation period 
obtained from the photometric data is a guarantee for the 
correct phase matching of the (usually less precise) RV data.
Therefore, the pulsation period and its variations have been 
determined with the method of the $O-C$ diagram \citep{S05}. 

Visual and photographic observations have not been taken into
account in the present study of the pulsation period. 
Photoelectric data have been available from the 1950s.

In the case of Cepheids pulsating with such a low amplitude, 
the $O-C$ diagram constructed for the median brightness is more 
reliable than that based on the moments of photometric maxima 
\citep{Detal12}. Therefore we determined the accurate value
of the pulsation period by constructing an $O-C$ diagram for
the moments of median brightness (the mid-point between the
faintest and the brightest states) on the ascending branch of 
light curve since it is this phase when the brightness variations 
are steepest during the whole pulsational cycle.

All published photoelectric and CCD photometric observations of 
FN~Aql covering more than 60 years were analysed in a homogeneous 
manner to determine seasonal moments of the chosen light-curve 
feature.

\begin{table}
\caption{$O-C$ values of FN~Aql (see the description 
in Section~\ref{fn-period})}
\begin{tabular}{l@{\hskip2mm}r@{\hskip2mm}r@{\hskip2mm}c@{\hskip2mm}l}
\hline
\noalign{\vskip 0.2mm}
JD$_{\odot}$ & $E\ $ & $O-C$ & $W$ & Data source\\
2\,400\,000 + &&&\\
\noalign{\vskip 0.2mm}
\hline
\noalign{\vskip 0.2mm}
33114.2710 & $-$2081 & $-$0.2087 & 3 & \citet{E51} \\
35276.1080 & $-$1853 & $-$0.1601 & 2 & \citet{I61} \\
35361.4481 & $-$1844 & $-$0.1538 & 2 & \citet{Wetal58} \\
37210.4186 & $-$1649 & $-$0.0812 & 1 & \citet{Metal64} \\
37949.9954 & $-$1571 & $-$0.0636 & 1 & \citet{W66} \\
40860.7681 & $-$1264 & $-$0.1200 & 3 & \citet{P76} \\
41116.8382 & $-$1237 & $-$0.0512 & 1 & \citet{FMcN80} \\
42197.7639 & $-$1123 & $-$0.0197 & 3 & \citet{V77} \\
42795.0282 & $-$1060 & $-$0.0917 & 3 & \citet{Sz77} \\
42899.2979 & $-$1049 & $-$0.1187 & 2 & \citet{D77} \\
44425.9318 &  $-$888 & $-$0.0108 & 1 & \citet{B08}\\
44463.8061 &  $-$884 & $-$0.0626 & 3 & \citet{MB84} \\
44738.8200 &  $-$855 & $-$0.0131 & 3 & \citet{E85} \\
46616.1558 &  $-$657 & $-$0.0198 & 3 & \citet{B08}\\
47422.1118 &  $-$572 &  0.0063 & 3 & \citet{B08}\\
47753.9208 &  $-$537 & $-$0.0382 & 3 & \citet{B08}\\
48417.6241 &  $-$467 & $-$0.0418 & 3 & {\it Hipparcos} \citep{ESA97}\\
48512.4782 &  $-$457 & $-$0.0030 & 3 & \citet{B08}\\
48645.2018 &  $-$443 & $-$0.0208 & 3 & \citet{Betal97} \\
48872.8363 &  $-$419 &  0.0570 & 3 & \citet{B08}\\
48948.5754 &  $-$411 & $-$0.0561 & 3 & \citet{Aetal98} \\
49356.2942 &  $-$368 & $-$0.0430 & 3 & \citet{Aetal98} \\
49432.2135 &  $-$360 &  0.0241 & 2 & \citet{Betal97} \\
49583.9003 &  $-$344 &  0.0064 & 2 & \citet{B08}\\
49944.1826 &  $-$306 &  0.0093 & 3 & \citet{B08}\\
50313.9465 &  $-$267 & $-$0.0250 & 3 & \citet{B08}\\
50636.3877 &  $-$233 &  0.0442 & 3 & \citet{B08}\\
50996.7174 &  $-$195 &  0.0759 & 3 & \citet{B08}\\
51271.5893 &  $-$166 & $-$0.0166 & 3 & \citet{B08}\\
51650.9581 &  $-$126 &  0.0911 & 3 & \citet{B08}\\
52314.5382 &   $-$56 & $-$0.0357 & 3 & ASAS \citep{P02}\\
52371.4848 &   $-$50 &  0.0217 & 3 & \citet{B08}\\
52741.2297 &   $-$11 & $-$0.0130 & 3 & {\it INTEGRAL} OMC \\
52845.5274 &     0 & $-$0.0121 & 3 & ASAS \citep{P02}\\
53177.4218 &    35 &  0.0288 & 3 & ASAS \citep{P02}\\
53205.8192 &    38 & $-$0.0184 & 2 & {\it INTEGRAL} OMC \\
53566.0826 &    76 & $-$0.0530 & 3 & ASAS \citep{P02}\\
53660.9556 &    86 &  0.0047 & 3 & {\it INTEGRAL} OMC \\
53860.0315 &   107 & $-$0.0315 & 3 & ASAS \citep{P02}\\
54305.7052 &   154 &  0.0104 & 3 & ASAS \citep{P02}\\
54324.6305 &   156 & $-$0.0274 & 3 & {\it INTEGRAL} OMC \\
54637.5900 &   189 &  0.0417 & 3 & AAVSO \\
54675.4691 &   193 & $-$0.0053 & 3 & ASAS \citep{P02}\\
54732.3398 &   199 & $-$0.0238 & 3 & {\it INTEGRAL} OMC \\
54931.5010 &   220 &  0.0253 & 3 & {\it INTEGRAL} OMC \\
55026.2887 &   230 & $-$0.0022 & 3 & ASAS \citep{P02}\\
55301.2081 &   259 & $-$0.0472 & 3 & AAVSO \\
56116.6646 &   345 & $-$0.0021 & 3 & AAVSO \\
\noalign{\vskip 0.2mm}
\hline
\end{tabular}
\label{tab-fnaql-oc}
\end{table}

\begin{figure}
\includegraphics[height=42mm, angle=0]{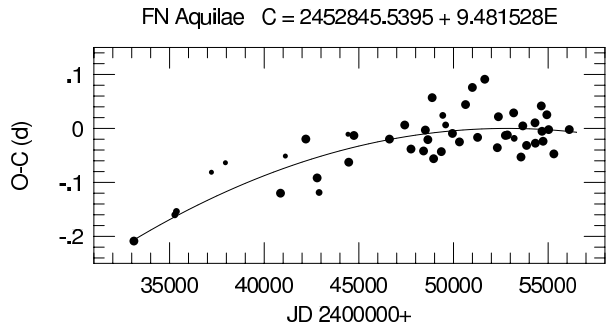}
\caption{$O-C$ diagram (for the median brightness on the 
ascending branch) of FN~Aql based on the values listed in 
Table~\ref{tab-fnaql-oc}. The pulsation period is subjected 
to a secular decrease.}
\label{fig-fnaql-oc}
\end{figure}

The relevant data listed in 
Table~\ref{tab-fnaql-oc} are as follows.\\
Column~1: heliocentric moment of the median brightness 
on the ascending branch;\\
Col.~2: epoch number, $E$, as calculated from 
equation~(\ref{fnaql-ephemeris}):
\vspace{-1mm}
\begin{equation}
C_{\rm med} = 2\,452\,845.5395 + 9.481\,528{\times}E - 4.77\times 10^{-8} E^2
\label{fnaql-ephemeris}
\end{equation}
\vspace{-3mm}
$\phantom{lmmmmmm}\pm0.0037\phantom{}\pm 0.000\,012\phantom{mm}\pm 0.75\times 10^{-8}$

\noindent (this ephemeris has been obtained by the weighted 
least-squares fit to the tabulated $O-C$ differences, where $E=0$ 
is selected arbitrarily and corresponds to the most reliable subset
of data);\\
\noindent Col.~3: the corresponding $O-C$ value as calculated 
from the constant and linear terms of equation~(\ref{fnaql-ephemeris});\\
Col.~4: weight assigned to the $O-C$ value (1, 2, or 3 
depending on the quality of the light curve leading to 
the given difference);\\
Col.~5: reference to the origin of data.

The plot of $O-C$ values shown in Fig.~\ref{fig-fnaql-oc}
can be fitted with a parabola implying a minute decrease 
($-0.3175$\,s\,yr$^{-1}$) in the pulsation period. Previous studies of
period variations of FN~Aql are also available in the literature.
\citet{Sz88} approximated the $O-C$ graph with a constant period and 
a light-time effect superimposed on it (see Sect.~\ref{fn-binarity}), 
while \citet{BP94} found a slightly increasing pulsation period. 
On the contrary, \citet{T98} included FN~Aql in the table of Cepheids 
with decreasing period. The value quoted by Turner was taken from the 
unpublished study carried out by Duncan (1991), and this value 
is smaller by 10\% than the period shortening derived in this paper.

\subsection{Binarity of FN~Aql}
\label{fn-binarity}

\begin{figure}
\includegraphics[height=55mm, angle=0]{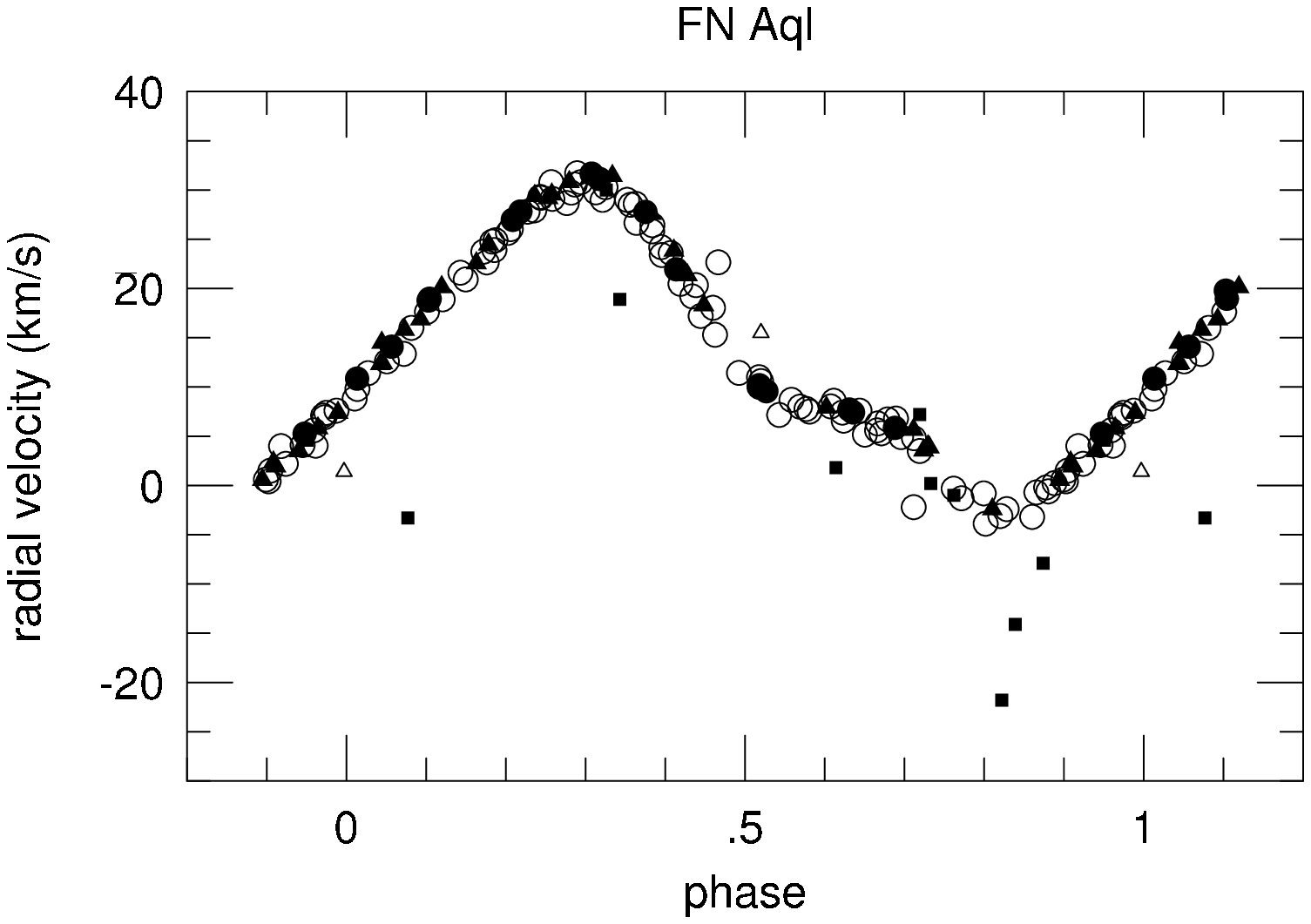}
\caption{Merged RV phase curve of FN~Aql. The filled circles 
represent our new data, open circles denote values obtained 
by Gorynya et~al. (\citeyear{Getal98}), filled triangles are 
those obtained by \citet{Betal05}, open triangles  represent 
values published by \citet{Betal88}, and Joy's (\citeyear{J37}) 
data are marked as black squares.
The RV data have been folded on the pulsation period of
9.481528 d, and the phase shifts due to the parabolic $O-C$
graph (see Fig.~\ref{fig-fnaql-oc}) have been applied for.
}
\label{fig-fnaql-vrad}
\end{figure}

\begin{figure}
\includegraphics[height=42mm, angle=0]{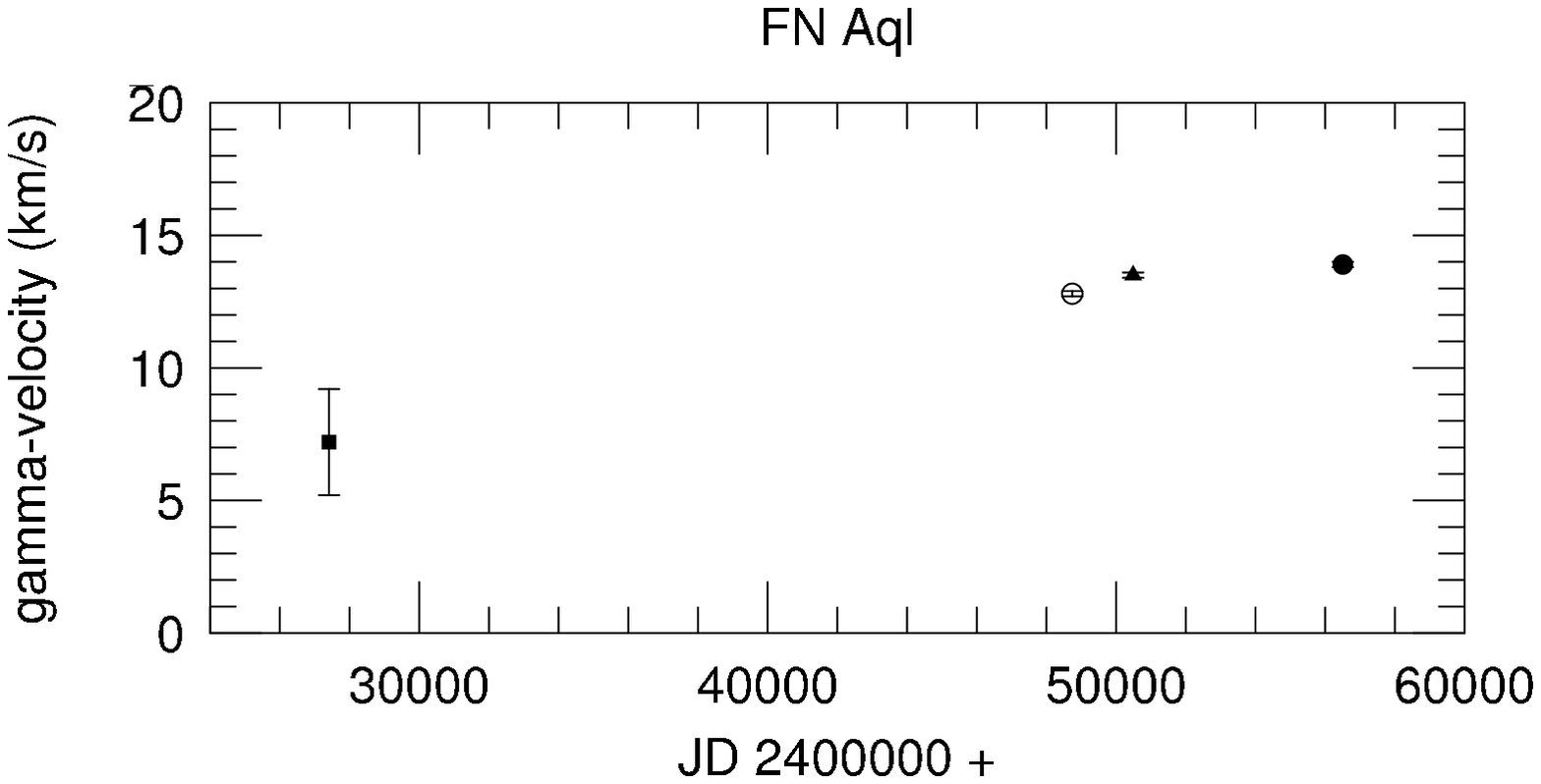}
\caption{Temporal variation in the $\gamma$-velocity of FN~Aql.
The symbols for the different data sets are the same as
in Fig.~\ref{fig-fnaql-vrad}.}
\label{fig-fnaql-vgamma}
\end{figure}

\begin{table}
\caption{$\gamma$-velocities of FN~Aql}
\begin{tabular}{lcccl}
\hline
\noalign{\vskip 0.2mm}
Mid-JD & $v_{\gamma}$ & $\sigma$ & $N$ & Data source \\
2\,400\,000+ & (km\,s$^{-1}$)& (km\,s$^{-1}$)& & \\
\noalign{\vskip 0.2mm}
\hline
\noalign{\vskip 0.2mm}
27412 &  +7.2& 2.0 & 11 & \citet{J37}\\
44916 &  uncertain & -- & 2 & \citet{Betal88}\\
48741 & +12.8& 0.1 & 94 & \citet{Getal98}\\
50488 & +13.5& 0.1 & 31 & \citet{Betal05}\\
56506 & +13.9& 0.1 & 17 & Present paper\\
\noalign{\vskip 0.2mm}
\hline
\end{tabular}
\label{tab-fnaql-vgamma}
\end{table}
 
A companion can have an observable effect on the colour,
amplitude and other phenomenological properties of a Cepheid
\citep{Sz03a}, even if the two stars are not physically bound
and only form optical pairs. Based on their multicolour 
photometry, \citet{D77} and \citet{P78} suspected a blue
companion but two other efficiently used photometric criteria 
did not indicate any hot companion \citep{M77,MF80}. Based on 
a UV spectrum obtained with the {\it IUE} satellite, \citet{Eetal90} 
excluded the presence of a companion hotter than A1V. 

Another piece of evidence of binarity is the light-time effect 
in the $O-C$ diagram of a pulsating variable. Though \citet{Sz88} 
suspected the presence of a light-time effect in the $O-C$
diagram of FN~Aql, no wave-like pattern is discernible
in the updated $O-C$ diagram in Fig.~\ref{fig-fnaql-oc}. The
suspected light-time effect was mainly due to the inclusion of
less reliable photographic data in the analysis of the pulsation
period by \citet{Sz88}. 

Nevertheless, there are some peculiarities in the behaviour of
FN~Aql. On the one hand, \citet{Uetal01} found a strong carbon 
deficit and a minor nitrogen overabundance in the spectrum of 
FN~Aql which cannot be explained by the canonical models of 
stellar evolution. On the other hand, \citet{Hetal08} pointed
out peculiar photometric properties of FN~Aql originally 
observed in the $U-B$ versus $B-V$ diagnostic plane. They explained
the correlations between the various colour indices in terms of
temporal variations in the circumstellar dust extinction.
Fluctuations in the observable stellar wind can be a consequence
of the orbital motion in a binary system. 

In addition to our own observations, RV data of FN~Aql are 
available from the following sources: \citet{J37}, \citet{Betal88}, 
\citet{Getal98}, and \citet{Betal05}. Using the accurate pulsation 
period determined from the $O-C$ diagram 
(equation~\ref{fnaql-ephemeris}) and applying the proper phase 
correction (corresponding to the parabolic $O-C$
graph), the resulting RV phase diagram is shown in 
Fig.~\ref{fig-fnaql-vrad}. Phase zero is arbitrarily chosen
at JD\,2\,400\,000. The meaning of the different symbols is
explained in Fig.~\ref{fig-fnaql-vrad}. It is seen that Joy's data 
are systematically more negative than the more recent RV 
measurements. Though the seminal paper by \citet{J37} reflects 
the observational technique and precision of the 1930s, 
and his method of deriving the RV value might be different 
from ours, Joy's data are free from any systematic errors in 
spite of their limited accuracy as documented by \citet{Sz96}.

The $\gamma$-velocities derived from four individual series of
RV observations are listed in Table~\ref{tab-fnaql-vgamma}
where the standard error of the $\gamma$-velocity and the number
of the data in the given series ($N$) is also listed. These 
data are also plotted in Fig.~\ref{fig-fnaql-vgamma}. It is 
clear that slight variations in the $\gamma$-velocity continued 
in the last two decades and the pattern of data in the diagram 
implies an orbital period of several decades for this SB system.


\section{Results for V1344~Aquilae}
\label{v1344aql}

\subsection{Accurate value of the pulsation period}
\label{v1344-period}

V1344~Aql pulsates in the first overtone mode, therefore it has 
a small pulsational amplitude and nearly sinusoidal light and 
velocity curves. The brightness variability of V1344~Aql was 
revealed by \citet{KSz79}. Curiously enough, this star was 
originally chosen as the check star for the photoelectric
observations of the Cepheid FN~Aql, in view of their similar
brightness and colour, as well as angular proximity. 

All published optical photometric observations of V1344~Aql 
covering 40 years were analysed in a homogeneous manner to 
determine seasonal moments of the median brightness on the
ascending branch of the light curve, similarly to the case
of FN~Aql. The $O-C$ diagram has been constructed from the 
$V$-band data (or nearest to this visual band).

The relevant data for constructing the $O-C$ graph are listed 
in Table~\ref{tab-v1344aql-oc} whose structure is similar to 
Table~\ref{tab-fnaql-oc}. The ephemeris of the $O-C$ residuals 
is:
\vspace{-1mm}
\begin{equation}
C_{\rm med} = 2\,450\,312.6665 + 7.476\,744{\times}E + 0.90 \times 10^{-7} E^2
\label{v1344aql-ephemeris}
\end{equation}
\vspace{-3mm}
$\phantom{lmmmmmm}\pm0.0106\phantom{}\pm 0.000\,017 \phantom{mm}\pm 0.32 \times 10^{-7} E^2$

\noindent as obtained from the second-order least-squares fit to the
heliocentric moments of the median brightness on the ascending branch.
The tabulated $O-C$ residuals have been obtained by using
the constant and linear terms of equation~(\ref{v1344aql-ephemeris}).

\begin{figure}
\includegraphics[height=42mm, angle=0]{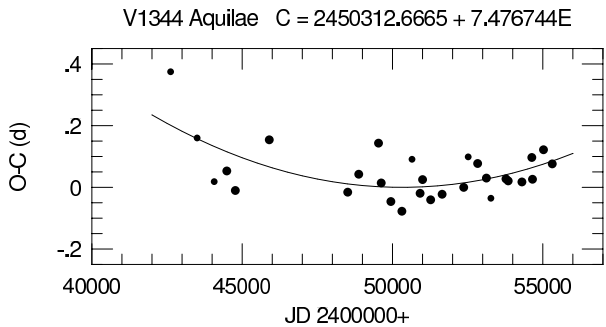}
\caption{$O-C$ diagram (for the median brightness on the 
ascending branch) of V1344~Aql based on the values
listed in Table~\ref{tab-v1344aql-oc}. The pattern of the graph
indicates that the pulsation period of V1344~Aql has been 
continuously increasing.}
\label{fig-v1344aql-oc}
\end{figure}

The plot of $O-C$ values shown in Fig.~\ref{fig-v1344aql-oc}
can be approximated with a parabola implying a minute increase 
in the pulsation period.

\begin{table}
\caption{$O-C$ values of V1344~Aql (see the description 
in Section~\ref{v1344-period})}
\begin{tabular}{l@{\hskip2mm}r@{\hskip2mm}r@{\hskip2mm}c@{\hskip2mm}l}
\hline
\noalign{\vskip 0.2mm}
JD$_{\odot}$ & $E\ $ & $O-C$ & $W$ & Data source\\
2\,400\,000 + &&&\\
\noalign{\vskip 0.2mm}
\hline
\noalign{\vskip 0.2mm}
42619.4717 & $-$1029 &  0.3748 & 2 & \citet{KSz79}\\
43501.5128 &  $-$911 &  0.1601 & 2 & \citet{KSz79} \\
44069.6038 &  $-$835 &  0.0185 & 2 & \citet{KSz79} \\
44488.3360 &  $-$779 &  0.0531 & 3 & \citet{FG81} \\
44772.3890 &  $-$741 & $-$0.0102 & 3 & \citet{A84} \\
45901.5416 &  $-$590 & +0.1541 & 3 &  \citet{E85}\\
48510.7555 &  $-$241 & $-$0.0157 & 3 & \citet{B08}\\
48877.1742 &  $-$192 &  0.0425 & 3 & \citet{B08}\\
49535.2284 &  $-$104 &  0.0971 & 3 & \citet{B08}\\
49624.8203 &   $-$92 &  0.0142 & 3 & \citet{B08}\\
49946.2600 &   $-$49 & $-$0.0460 & 3 & \citet{B08}\\
50312.5890 &     0  &$-$0.0775 & 3 & \citet{B08}\\
50649.2110 &    45  &   0.0910 & 2 & \citet{B08}\\
50918.2635 &    81  &$-$0.0193 & 3 & \citet{IV00}\\
51000.5520 &    92  &   0.0251 & 3 & \citet{B08}\\
51269.6493 &   128  &$-$0.0404 & 3 & \citet{B08}\\
51650.9813 &   179  &$-$0.0224 & 3 & \citet{B08}\\
52368.7710 &   275  &$-$0.0001 & 3 & \citet{B08}\\
52518.4048 &   295  &   0.0988 & 2 & ASAS \citep{P02}\\
52832.4065 &   337  &   0.0773 & 3 & ASAS \citep{P02}\\
53123.9522 &   376  &   0.0300 & 3 & ASAS \citep{P02}\\
53273.4217 &   396  &$-$0.0354 & 2 & ASAS \citep{P02}\\
53774.4264 &   463  &   0.0274 & 3 & ASAS \citep{P02}\\
53849.1873 &   473  &   0.0209 & 3 & ASAS \citep{P02}\\
54305.2653 &   534  &   0.0175 & 3 & ASAS \citep{P02}\\
54634.3213 &   578  &   0.0968 & 3 & AAVSO \\
54656.6810 &   581  &   0.0262 & 3 & ASAS \citep{P02}\\
55023.1371 &   630  &   0.1219 & 3 & ASAS \citep{P02}\\
55314.6843 &   669  &   0.0761 & 3 & AAVSO \\
\noalign{\vskip 0.2mm}
\hline
\end{tabular}
\label{tab-v1344aql-oc}
\end{table}

\subsection{Binarity of V1344~Aquilae}
\label{v1344-binarity}

Due to the belated discovery of brightness variability in V1344~Aql,
this Cepheid has not been on the target list of most of the projects
aimed at revealing companions to Cepheids from photometric data. 
The wavelength dependence of the photometric amplitudes studied by
\citet{KSz09}, however, hints at the presence of a red companion.

RV data have been available from two consecutive years: from 1980 
by \citet{B81} and from 1981 by \citet{A84}. Additional spectroscopy 
of V1344~Aql was performed by \citet{LL11} but they concentrated on 
the chemical composition of their numerous target Cepheids without
determining RV values from the spectra. They obtained an atmospheric 
iron abundance of [Fe/H]= 0.15 for V1344~Aql.

\begin{figure}
\includegraphics[height=55mm, angle=0]{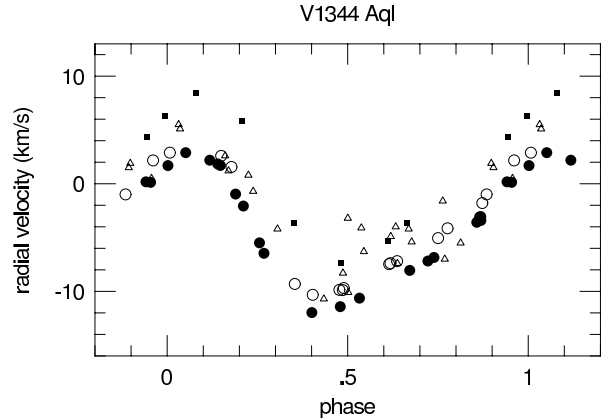}
\caption{Merged RV phase curve of V1344~Aql. The filled and open 
circles represent our new data obtained in 2012 and 2013, 
respectively, triangles denote Balona's (\citeyear{B81}) 
data, while Arellano Ferro's (\citeyear{A84}) data are marked as 
black squares.}
\label{fig-v1344aql-vrad}
\end{figure}

\begin{figure}
\includegraphics[height=42mm, angle=0]{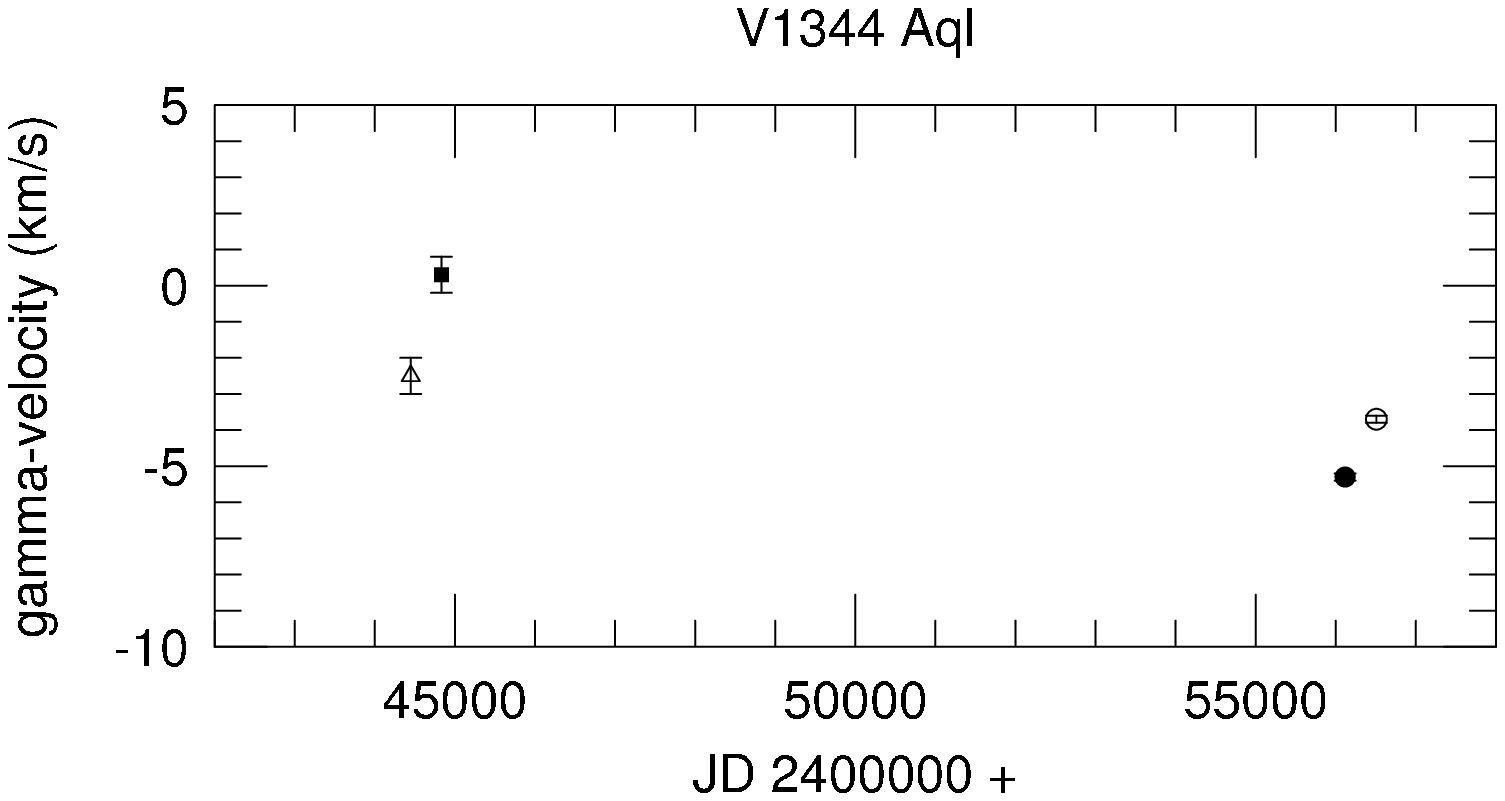}
\caption{Temporal variation in the $\gamma$-velocity of V1344~Aql.
The symbols for the different data sets are the same as
in Fig.~\ref{fig-v1344aql-vrad}.}
\label{fig-v1344aql-vgamma}
\end{figure}

\begin{table}
\caption{$\gamma$-velocities of V1344~Aql}
\begin{tabular}{lcccl}
\hline
\noalign{\vskip 0.2mm}
Mid-JD & $v_{\gamma}$ & $\sigma$ & $N$ &Data source \\
2\,400\,000+ & (km\,s$^{-1}$)& (km\,s$^{-1}$) & & \\
\noalign{\vskip 0.2mm}
\hline
\noalign{\vskip 0.2mm}
44449 & $-$2.5& 0.5 & 24 & \citet{B81}\\
44832 & +0.3& 0.5 & 8 & \citet{A84}\\
56117 & $-$5.3& 0.1 & 22 & Present paper\\
56507 & $-$3.7& 0.1 & 16 & Present paper\\
\noalign{\vskip 0.2mm}
\hline
\end{tabular}
\label{tab-v1344aql-vgamma}
\end{table}

The two earlier RV data series for V1344~Aql \citep{B81,A84} 
already imply a slight shift between the annual mean velocity.
(When constructing the phase curve from these data, it became 
obvious that there might be a misprint in the list of Arellano 
Ferro's data: instead of $-$14.31 km\,s$^{-1}$, the correct value 
should be +4.31 km\,s$^{-1}$ at JD\,2\,444\,829.762.)

Supplemented with our new data, the merged phase diagram of all 
RV observations is plotted in Fig.~\ref{fig-v1344aql-vrad}.
In order to reach correct phasing of data obtained in widely
different epochs, the RV data have been folded on the pulsation
period given in the ephemeris in equation~(\ref{v1344aql-ephemeris}). 
The zero phase has been arbitrarily chosen at JD\,2\,400\,000. 
The data in Fig.~\ref{fig-v1344aql-vrad} clearly show a vertical 
shift between the mean RV values exceeding 2\,km\,s$^{-1}$ within 
one year, referring to the membership of V1344~Aql in a SB system.

Table~\ref{tab-v1344aql-vgamma} summarizes the pulsation averaged 
mean RV values. Variability in the $\gamma$-velocity is visualized
in Fig.~\ref{fig-v1344aql-vgamma}. The pattern of the data points
in this figure implies an orbital period of several hundred days
which is rather short among the SB Cepheids.

\section{Conclusions}
\label{concl}

We pointed out that the classical Cepheids FN~Aql and
V1344~Aql have a variable $\gamma$-velocity which implies 
their membership in SB systems. The available RV data are 
insufficient to determine the orbital period and other 
elements of the orbit. However, the temporal variations in the 
$\gamma$-velocity indicate an orbital period as long as 
several decades for the FN~Aql system and a rather short 
orbital period for the SB system involving V1344~Aql.

The value of the orbital period for SB systems with a 
Cepheid component is often unknown: according to the on-line 
data base \citep{Sz03b} the orbital period has been determined 
for about 20\% of the known SB Cepheids. For most of them, 
the known orbital period exceeds a thousand days.

Our finding confirms the previous statement by \citet{Sz03b} 
about the high percentage of binaries among classical Cepheids 
and the observational selection effect hindering the discovery 
of new cases. Moreover, another statistical bias is apparent
from the list of Galactic Cepheids known in binary systems
consisting of 165 items: only 40\% of the Cepheids have a 
positive declination, indicating that the northern sky has not
been investigated satisfactorily.

Regular monitoring of the radial velocities of a large
number of Cepheids will be instrumental in finding 
more long-period spectroscopic binaries among Cepheids. 
RV data to be obtained with the {\it Gaia}
astrometric space probe (launched on 19 Dec 2013)
will certainly result in revealing many new spectroscopic
binaries among Cepheids brighter than 13--14 mag.

In principle, the orbital motion gives rise to a light-time 
effect in the $O-C$ diagram of pulsating stars. In our case, 
however, its amplitude is either too small (for V1344~Aql, the
effect is of the order of a  thousandth of a day), or it cannot 
be revealed yet because of the length of the orbital period
(several decades in the case of FN~Aql). Both Cepheids show
secular variations in the pulsation period (see the $O-C$ diagrams 
in Figs~\ref{fig-fnaql-oc} and \ref{fig-v1344aql-oc}, respectively).
In addition to these changes of evolutionary origin, the recently 
discovered period jitter in classical Cepheids \citep{Detal12} 
is also against revealing a very low amplitude light-time effect 
in the $O-C$ diagrams of Cepheids.

\section*{Acknowledgments} 

This project has been supported by the  
ESTEC Contract no.\,4000106398/12/NL/KML, the Hungarian OTKA 
Grant K83790, the European Community's Seventh Framework 
Programme (FP7/2007-2013) under grant agreement no.~269194,
as well as City of Szombathely under agreement no.~S-11-1027,
the `Lend\"ulet-2009' Young Researchers Programme of the 
Hungarian Academy of Sciences, and the Mag Zrt. MB08C 81013
project. 
The {\it INTEGRAL\/} photometric data, pre-processed by 
ISDC, have been retrieved from the OMC Archive at CAB (INTA-CSIC). 
The photometric contribution of AAVSO observers and the
clarifying remarks by the referee are gratefully acknowledged.

\bsp

\label{lastpage}


\begin{thebibliography}{99}

\bibitem[\protect\citeauthoryear{Arellano Ferro}{1984}]{A84}
Arellano Ferro A. 1984, MNRAS, 209, 481
\bibitem[\protect\citeauthoryear{Arellano Ferro et al.}{1998}]{Aetal98}
Arellano Ferro A., Rojo Arellano E., Gonzalez-Bedolla S., Rosenzweig, P.
1998, ApJS, 117, 167
\bibitem[\protect\citeauthoryear{Balona}{1981}]{B81}
Balona L.~A. 1981, The Observatory, 101, 205
\bibitem[\protect\citeauthoryear{Barnes et~al.}{1988}]{Betal88}
Barnes T.~G., III, Moffett T.~J., Slovak M.~H. 1988, ApJS, 66, 43
\bibitem[\protect\citeauthoryear{Barnes et~al.}{1997}]{Betal97}
Barnes T.~G., III, Fernley J.~A., Frueh M.~L., Navas J.~G., 
Moffett T.~J., Skillen I. 1997, PASP, 109, 645
\bibitem[\protect\citeauthoryear{Barnes et~al.}{2005}]{Betal05}
Barnes T.~G., III, Jeffery E.~J., Montemayor T.~J., Skillen I. 
2005, ApJS, 156, 227
\bibitem[\protect\citeauthoryear{Berdnikov}{2008}]{B08}Berdnikov
L.~N. 2008, VizieR On-line Data Catalog: II/285
\bibitem[\protect\citeauthoryear{Berdnikov \& Pastukhova}{1994}]
{BP94}Berdnikov L.~N., Pastukhova E.~N. 1994, Astron. Lett., 20, 479
\bibitem[\protect\citeauthoryear{Bla\v{z}ko}{1929}]{B29}Bla\v{z}ko S.
1929, Astron. Nachr., 236, 279
\bibitem[\protect\citeauthoryear{Cs\'ak et al.}{2014}]{Csetal14}
Cs\'ak B., Kov\'acs, Szab\'o Gy.~M., Kiss L.~L., D\'ozsa \'A.,
S\'odor \'A., Jankovics I. 2014, Contrib. Astron. Obs. Skalnat\'e Pleso,
43, 189
\bibitem[\protect\citeauthoryear{Dean}{1977}]{D77}Dean J.~F. 1977, 
MNASSA, 36, 3
\bibitem[\protect\citeauthoryear{Derekas et~al.}{2012}]{Detal12}
Derekas A. et~al. 2012, MNRAS, 425, 1312
\bibitem[\protect\citeauthoryear{Eggen}{1985}]{E85}Eggen O.~J. 1985,
AJ, 90, 1297
\bibitem[\protect\citeauthoryear{Eggen}{1951}]{E51}Eggen O.~J. 1951,
ApJ, 113, 367
\bibitem[\protect\citeauthoryear{ESA}{1997}]{ESA97}ESA 1997, 
ESA SP-1200, The Hipparcos and Tycho Catalogues, ESA publ. Division,
Noordwijk
\bibitem[\protect\citeauthoryear{Evans}{1992}]{E92}
Evans N.~R. 1992, ApJ, 389, 657
\bibitem[\protect\citeauthoryear{Evans et~al.}{1990}]{Eetal90}
Evans N.~R., Szabados L., Udalska, J. 1990, PASP, 102, 981
\bibitem[\protect\citeauthoryear{Feltz \& McNamara}{1980}]{FMcN80}
Feltz K.~A., Jr., McNamara D.~H. 1980, PASP, 92, 609
\bibitem[\protect\citeauthoryear{Fernie \& Garrison}{1981}]{FG81}
Fernie J.~D., Garrison R.~F. 1981, PASP, 93, 330
\bibitem[\protect\citeauthoryear{Gorynya et~al.}{1998}]{Getal98}
Gorynya N.~A., Samus' N.~N., Sachkov M.~E., Rastorguev A.~S., 
Glushkova E.~V., Antipin S.~V. 1998, Astron. Lett., 24, 815	
\bibitem[\protect\citeauthoryear{Hrudkov\'a}{2006}]{Hrudkova2006}
Hrudkov\'a M. 2006, in Safrankova J., Pavlu~J., eds, WDS 06, Proc. 
of Contributed Papers: Part III -- Physics. Matfyzpress, Prague, 18
\bibitem[\protect\citeauthoryear{Hurley et~al.}{2008}]{Hetal08}
Hurley M., Madore B.~F., Freedman W.~L. 2008, AJ, 135, 2217
\bibitem[\protect\citeauthoryear{Ignatova \& Vozyakova}{2000}]{IV00}
Ignatova V.~V., Vozyakova O.~V. 2000, Astron. Astrophys. Trans., 19, 133
\bibitem[\protect\citeauthoryear{Irwin}{1961}]{I61}Irwin J.~B. 1961,
ApJS, 6, 253
\bibitem[\protect\citeauthoryear{Joy}{1937}]{J37}
Joy A.~H. 1937, ApJ, 86, 363
\bibitem[\protect\citeauthoryear{Klagyivik \& Szabados}{2009}]{KSz09}
Klagyivik P., Szabados L. 2009, A\&A, 504, 959
\bibitem[\protect\citeauthoryear{Kovacs \& Szabados}{1979}]{KSz79}
Kovacs G., Szabados L. 1979, Inf. Bull. Var. Stars, 1719, 1
\bibitem[\protect\citeauthoryear{Luck \& Lambert}{2011}]{LL11}
Luck R.~E., Lambert D.~L. 2011, AJ, 142, 136
\bibitem[\protect\citeauthoryear{Madore}{1977}]{M77}Madore B.~F.
1977, MNRAS, 178, 505
\bibitem[\protect\citeauthoryear{Madore \& Fernie}{1980}]{MF80}Madore
B.~F., Fernie J.~D. 1980, PASP, 92, 315
\bibitem[\protect\citeauthoryear{Mitchell et~al.}{1964}]{Metal64}
Mitchell R.~I., Iriarte B., Steinmetz D., Johnson H.~L. 1964,
Bol. Obs. Tonantzintla Tacubaya, 3, 153
\bibitem[\protect\citeauthoryear{Moffett \& Barnes}{1984}]{MB84}
Moffett T.~J., Barnes T.~G. 1984, ApJS, 55, 389
\bibitem[\protect\citeauthoryear{Munari et al.}{2005}]{Munari2005}
Munari U., Sordo R., Castelli F., Zwitter T. 2005, A\&A, 442, 1127 
\bibitem[\protect\citeauthoryear{Pel}{1976}]{P76}Pel J.~W. 1976,
A\&AS, 24, 413 
\bibitem[\protect\citeauthoryear{Pel}{1978}]{P78}Pel J.~W. 1978,
A\&A, 62, 75 
\bibitem[\protect\citeauthoryear{Pojmanski}{2002}]{P02}Pojmanski G.
2002, Acta Astron., 52, 397
\bibitem[\protect\citeauthoryear{Prager}{1931}]{P31}Prager R. 1931,
Astron. Nachr., 243, 359
\bibitem[\protect\citeauthoryear{Sterken}{2005}]{S05}
Sterken C. 2005, in Sterken C. ed., ASP Conf. Ser. Vol.~335, The 
Light-Time Effect in Astrophysics. Astron. Soc. Pac., San Francisco, p.\,3
\bibitem[\protect\citeauthoryear{Sza\-bados}{1977}]{Sz77}Szabados L. 
1977, Mitt. Sternw. ung. Akad. Wiss., Budapest, No.\,70
\bibitem[\protect\citeauthoryear{Sza\-bados}{1988}]{Sz88}Szabados L. 
1988, PASP, 100, 589
\bibitem[\protect\citeauthoryear{Sza\-bados}{1996}]{Sz96}Szabados L.
1996, A\&A, 311, 189
\bibitem[\protect\citeauthoryear{Sza\-bados}{2003a}]{Sz03a}Szabados L. 
2003b, in Recent Res. Devel. Astron. \& Astrophys., 1, 787
\bibitem[\protect\citeauthoryear{Sza\-bados}{2003b}]{Sz03b}Szabados L. 
2003a, Inf. Bull. Var. Stars, 5394 
\bibitem[\protect\citeauthoryear{Szabados \& Klagyivik}{2012}]{SzK12}
Szabados L., Klagyivik P. 2012, Ap\&SS, 341, 99
\bibitem[\protect\citeauthoryear{Thizy \& Cochard}{2011}]{Thizy2011}
Thizy O., Cochard F. 2011, in Wade~G., Meynet~G., Peters~G., Neiner~C.,
eds, Proc. IAU Symp. 272, Active OB Stars: Structure, Evolution, Mass 
Loss, and Critical Limits. Cambridge Univ. Press, Cambridge, p.\,282
\bibitem[\protect\citeauthoryear{Turner}{1998}]{T98}Turner D.~G.
1998, J. Am. Assoc. Var. Star Obser., 26, 101
\bibitem[\protect\citeauthoryear{Usenko et~al.}{2001}]{Uetal01}
Usenko I.~A., Kovtyukh V.~V., Klochkova V.~G. 2001, A\&A, 377, 156
\bibitem[\protect\citeauthoryear{Vasil'yanovskaya}{1977}]{V77}
Vasil'yanovskaya O.~P. 1977, Perem. Zvezdy, 20, 467
\bibitem[\protect\citeauthoryear{Walraven et~al.}{1958}]{Wetal58}
Walraven Th., Muller A.~B., Oosterhoff P.~T. 1958, Bull. Astron. Inst. 
Neth., 14, 81
\bibitem[\protect\citeauthoryear{Williams}{1966}]{W66}
Williams, J.~A. 1966, AJ, 71, 615
\end{thebibliography}
\end{document}